\documentclass{ctr_summer}

\usepackage{ctrfont}
\usepackage{natbib}
\usepackage{undertilde}
\usepackage{color}

\usepackage{graphicx}
\usepackage{psfrag}
\usepackage{epsfig}
\usepackage{amsmath,rotating}
\usepackage{dashrule}
\usepackage{undertilde}

\usepackage{url}

\usepackage{color}

\title{Large eddy simulation of ocean mesoscale eddies}

\shorttitle{LES of ocean mesoscale eddies}

\author{P. Perezhogin\footnote[1]{Courant Institute of Mathematical Sciences, New York University}, A. Balakrishna \and R. Agrawal}

\shortauthor{Perezhogin, Balakrishna \& Agrawal}

\begin{document}

\setcounter{page}{1}
\maketitle
Mesoscale eddies produce lateral (2D) fluxes that need to be parameterized in eddy-permitting ($1/4^\circ$) global ocean models due to insufficient horizontal resolution.
Here, we systematically apply methods from the 3D LES community to parameterize lateral vorticity fluxes produced by mesoscale eddies leveraging an explicit filtering approach together with a dynamic procedure. The developed subfilter closure is implemented into the GFDL MOM6 ocean model and is evaluated in an idealized double-gyre configuration, both \emph{a-priori} and \emph{a-posteriori}. For sufficiently resolved grids, the LES simulations converge to the filtered high-resolution data. 
However, limitations in the proposed closure are observed when the filter scale approaches the energy-containing scales: the \emph{a-priori} performance drops and \emph{a-posteriori} experiments fail to converge to the filtered high-resolution data. Nevertheless, the proposed closure is accurate in predicting the mean flow in \emph{a-posteriori} simulations at all resolutions considered ($1/2^{\circ}-1/8^{\circ}$). Finally, we propose parameterizing the thickness fluxes using a Bardina model which further improves simulations at the coarsest resolutions ($1/2^\circ-1/3^\circ$).
\\

\hrule
\section{Introduction}
Traditional parameterizations of mesoscale eddies attempt to close the energy budget by parameterizing dissipation and backscatter of kinetic energy (KE) \citep{jansen2014parameterizing} and removal of the potential energy \citep{gent1990isopycnal}. However, these parameterizations are commonly tuned to maximize \emph{a-posteriori} performance, while their \emph{a-priori} accuracy and connection to high-resolution data are unknown. 
In recent years, there has been a substantial effort to develop mesoscale parameterizations with high \emph{a-priori} accuracy by applying the large eddy simulation (LES) paradigm \citep{fox2008can,khani2023gradient} or machine learning methods \citep{zanna2020data}. However, as compared to studies of 3D turbulence, the assessment of these methods has been limited thus far. For instance, the dynamic model of \cite{germano1991dynamic} has been evaluated in ocean models only once \citep{bachman2017scale}. \\


\noindent
Vorticity fluxes (or momentum fluxes) in quasi-2D fluids are responsible for the forward cascade of enstrophy and inverse cascade of energy. Thus, careful consideration of both dissipation and backscatter is required to produce accurate and numerically stable subfilter closures.
\cite{perezhogin2023subgrid} showed that the 3D LES closure of \cite{horiuti1997new} can be adapted for the prediction of vorticity fluxes in 2D decaying turbulence. In particular, the Reynolds stress, which is a part of the \cite{germano1986proposal} decomposition, is responsible for additional KE backscatter. This finding was further confirmed and exploited by \cite{perezhogin2023implementation} in an idealized GFDL MOM6 ocean model \citep{adcroft2019gfdl}. However, subfilter closures in that work were subject to substantial \emph{a-posteriori} tuning, and thus assessing their \emph{a-priori} accuracy is challenging. \\

\noindent  
In this light, we propose a methodology for parameterization of mesoscale eddies in the MOM6 ocean model that addresses a problem of \emph{a-posteriori} tuning without degrading \emph{a-priori} accuracy. We utilize two approaches adopted by the 3D LES community, namely the explicit filtering approach \citep{winckelmans2001explicit, bose2010grid} and the dynamic procedure of \cite{germano1991dynamic}. 
Following the explicit filtering approach, we assume that the LES filter width $\overline{\Delta}$ is larger than the LES grid spacing $\Delta x$.
This improves our predictions of the subfilter flux by allowing a partial inversion of the LES filter \citep{carati2001modelling}. Additionally, we suppress the numerical discretization errors by increasing the filter-to-grid width ratio ($\mathrm{FGR}=\overline{\Delta}/\Delta x$). Finally, the dynamic procedure allows us to accurately estimate the free parameters of the parameterization from the resolved flow only. Overall, our approach permits a parameterization with good \emph{a-priori} accuracy and contains a single tunable parameter FGR, which is adjusted to maximize the \emph{a-posteriori} performance. \\ 

\noindent
The rest of this report is organized as follows. Section 2 describes the MOM6 ocean model, the subfilter closure of vorticity fluxes \citep{perezhogin2023subgrid} and a newly proposed closure of thickness fluxes. In Section 3, we show \emph{a-priori} and \emph{a-posteriori} results and compare them to high-resolution filtered data. Finally, concluding remarks are made in Section 4.


\section{Methods}
%


Here, we attempt to improve the dynamical core of the MOM6 ocean model, which solves the stacked shallow water equations, with improved LES subfilter closures. 

\subsection{Double-Gyre configuration}
The stacked (i.e., multilayer) shallow water equations in the vector-invariant form are given as
\begin{gather}
\partial_t u^k - (\omega^k + f) v^k + \partial_x (K^k + M^k) = \mathcal{F}^k_x, \label{eq:f0} \\
\partial_t v^k + (\omega^k + f)  u^k + \partial_y (K^k + M^k) = \mathcal{F}^k_y, \label{eq:f1} \\
\partial_t h^k + \partial_x (u^kh^k) + \partial_y (v^kh^k) = 0, \label{eq:f2}
\end{gather}
where $k$ is the index of the fluid layer, $u^k$ and $v^k$ are zonal and meridional velocities, and $h^k$ is the thickness of the fluid layer. Further, $\omega^k=\partial_x v^k - \partial_y u^k$ is the relative vorticity, $K^k=1/2((u^k)^2+(v^k)^2)$ is the KE per unit mass, and $M^k$ is the Montgomery potential (pressure anomaly). Also, $f$ is the Coriolis parameter and $\mathcal{F}^k$ is an additional forcing and dissipation. Note that there is no horizontal molecular viscosity in these equations and thus the small-scale dissipation is provided by the subgrid parameterization only.

\begin{figure}
\centering{\includegraphics[width=1\textwidth]{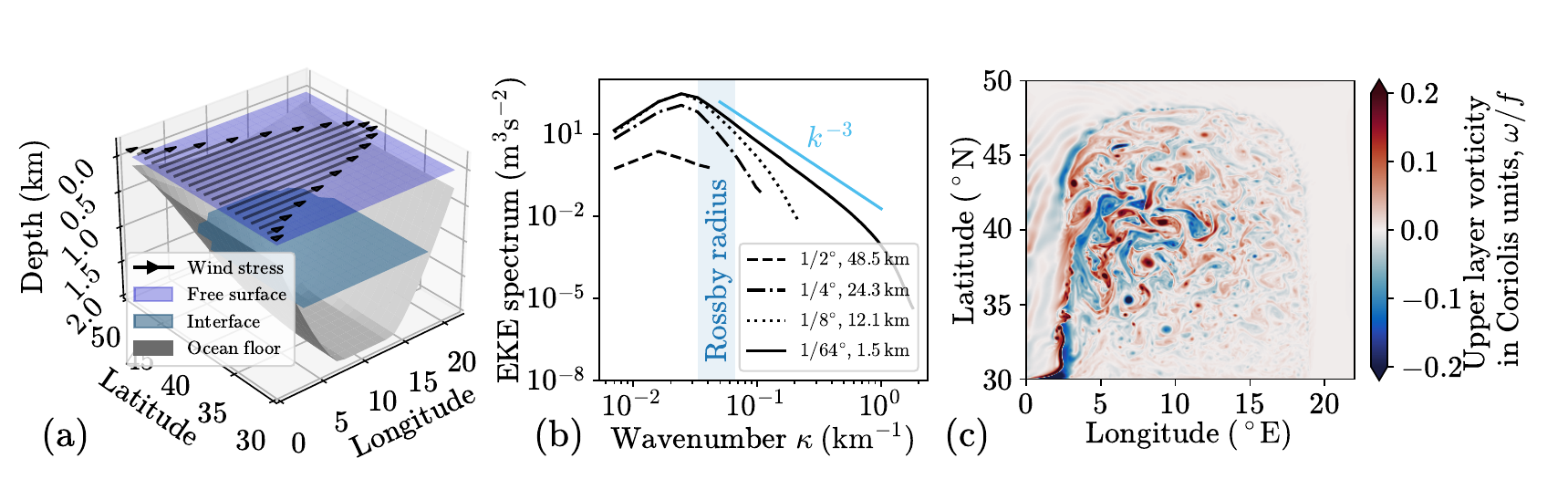}}
\caption{Idealized GFDL MOM6 ocean model in a double-gyre configuration parameterized with biharmonic Smagorinsky model ($C_S=0.06$): (a) computational domain, (b) spatial spectrum of eddy kinetic energy (EKE) in a box $(5^\circ$$-$$15^\circ\mathrm{E})$$~\times~$$(35^\circ$$-$$45^\circ\mathrm{N})$ in the upper fluid layer, (c) snapshot in the high-resolution model ($1/64^\circ$).} \label{Fig1}
\end{figure}

We consider an idealized GFDL MOM6 ocean model in a classical double-gyre configuration described in \cite{perezhogin2023implementation} and shown in Figure \ref{Fig1}. Two fluid layers are governed by the stacked shallow water equations (Eqs. \eqref{eq:f0}-\eqref{eq:f2}), with $\mathcal{F}^k$ given by a wind stress applied at the surface and a drag force applied at the ocean floor. The model domain is on a sphere with a spatial extent of $(0^\circ$$-$$22^\circ\mathrm{E})$$~\times~$$(30^\circ$$-$$50^\circ\mathrm{N})$ which is approximately $2000~\mathrm{km}$$~\times~$$2000~ \mathrm{km}$. We run experiments for 20 years of model time and use the last 10 years for computation of statistics to isolate from the transient development.

The high-resolution simulation at resolution $1/64^\circ$ resolves the mesoscale eddies with 10 grid points, as its grid spacing ($1.5$~km) is 10 times smaller than the first baroclinic Rossby radius ($15-30$~km), the scale at which mesoscale eddies are generated (see blue shading in Figure \ref{Fig1}(b)). In contrast, ocean models with the coarsest resolutions considered  ($1/2^\circ$ and $1/4^\circ$) have an insufficient number of grid points per Rossby radius, leading to insufficient eddy energy across all spatial scales (Figure \ref{Fig1}(b)); here eddies are defined as deviations from the 10-years average. In considered baseline experiments, we parameterize the subgrid momentum fluxes with the biharmonic Smagorinsky model having a coefficient of $C_S=0.06$, which is a default value in the global ocean model \citep{ adcroft2019gfdl}. Smaller values of $C_S$ are possible but will lead to the appearance of numerical noise in the vorticity field.  The mentioned lack of eddy energy is caused by the too-strong dissipation provided by the biharmonic Smagorinsky parameterization. The remainder of this report focuses on replacing this parameterization with alternative subgrid models to enhance the coarse-resolution ocean models.

\subsection{LES of stacked shallow water equations}
Equations \eqref{eq:f0}-\eqref{eq:f2} can be spatially filtered $\overline{(\cdot)}$ to arrive at
\begin{gather}
    \partial_t \overline{u} - (\overline{\omega} +f) \,\overline{v} + \partial_x (\overline{K}+\overline{M}) = \overline{\mathcal{F}_x} + \sigma_y, \label{eq:F1} \\
    \partial_t \overline{v} + (\overline{\omega} +f) \, \overline{u} + \partial_y (\overline{K}+\overline{M}) = \overline{\mathcal{F}_y} -\sigma_x, \label{eq:F2} \\
    \partial_t \overline{h} + \partial_x (\overline{u} \overline{h}) + \partial_y (\overline{v} \overline{h}) = - \partial_x \lambda_x - \partial_y \lambda_y, \label{eq:F3}
\end{gather}
where fluid layer index $k$ has been omitted for brevity. Further, 
\begin{equation}
 \sigma_j = \overline{u_j \omega} - \overline{u}_j \overline{\omega}    \hspace{10pt} \mathrm{and} \hspace{10pt} \lambda_j = \overline{u_j h} - \overline{u}_j \overline{h} \hspace{10pt} \mathrm{with} \hspace{10pt} j\in\{1,2\} 
\end{equation}
are the subfilter-scale (SFS) vorticity flux and subfilter thickness fluxes (i.e., mass fluxes), respectively.

The main effort of this work is to parameterize vorticity fluxes. Note that in the momentum equations (Eqs. \eqref{eq:F1}-\eqref{eq:F2}), we neglect the contribution of the subfilter KE because it does not affect flows with zero horizontal divergence \citep{marshall2010parameterization}. Our approach can be enhanced by parameterizing subfilter momentum fluxes rather than the vorticity fluxes. This adjustment would incorporate the local conservation law of momentum and naturally account for the horizontally divergent flows.  
However, extracting the momentum fluxes would require applying thickness-weighted averaging \citep{loose2023comparing}, also known as Favre averaging, which was previously adopted for LES of compressible flows by \cite{moin1991dynamic}.


\subsection{Vorticity closure} \label{sec:vorticity_closure}
Here, we consider the subfilter model of \cite{perezhogin2023subgrid}
\begin{equation}
    \overline{u_j \omega} - \overline{u}_j \overline{\omega}  \approx \overline{\overline{u}_j \overline{\omega}} - \overline{\overline{u}}_j \overline{\overline{\omega}} + C_S(\Delta x)^4  |\overline{S}| \frac{\partial \nabla^2 \overline{\omega}}{\partial x_j} + C_R \big(\overline{\overline{u_j'}~ \overline{\omega'}} - \overline{\overline{u_j'}}~ \overline{\overline{\omega'}} \big), \label{eq:closure}
\end{equation}
which consists of the Leonard flux, biharmonic Smagorinsky model, and the Reynolds-stress model, respectively, where $|\overline{S}|=\sqrt{(\partial_x \overline{u} - \partial_y \overline{v})^2 + (\partial_y \overline{u} + \partial_x \overline{v})^2}$ is the modulus of the strain-rate tensor, $\nabla^2=\partial_x^2 + \partial_y^2$ is the Laplace operator, and $\Delta x$ is the grid spacing of the coarse LES model. Note that for a quantity $\phi$, we define $\overline{\phi'} = \overline{\phi} - \overline{\overline{\phi}}$ and $\overline{\overline{\phi'}} = \overline{\overline{\phi}} - \overline{\overline{\overline{\phi}}}$. 

\subsubsection{Dynamic estimation of $C_S$ and $C_R$}

The subfilter closure (Eq. \eqref{eq:closure}) has two free parameters, $C_S$ and $C_R$, which can be estimated with the dynamic procedure \cite[see][appendix D]{perezhogin2023subgrid}. Herein, we describe only a modification of this dynamic procedure using a scale-similarity dynamic procedure \cite[see][section 4.2]{yuan2022dynamic}. In our work, we found that the \cite{yuan2022dynamic} dynamic procedure has several advantages as compared to \cite{germano1991dynamic}, specifically for the MOM6 ocean model: reduced computational cost, higher \emph{a-priori} accuracy, and lower sensitivity to the boundary artifacts. \\

\noindent
The dynamic procedure of \cite{yuan2022dynamic} can be further simplified following \cite{perezhogin2023subgrid} who assume that a test filter $\widehat{(\cdot)}$ is equal to the base filter, $\widehat{(\cdot)} \equiv \overline{(\cdot)}$, and thus the subfilter closure on the test-filter level is given in Eq. \eqref{eq:closure}. Then, substituting unfiltered fields ($u_j,\omega$) with filtered fields ($\overline{u}_j, \overline{\omega}$) in Eq. \eqref{eq:closure}, we obtain an analog of the \cite{germano1991dynamic} identity for our subfilter closure
\begin{equation}
    \overline{\overline{u}_j \overline{\omega}} - \overline{\overline{u}}_j \overline{\overline{\omega}} \approx
    \overline{\overline{\overline{u}}_j \overline{\overline{\omega}}} - \overline{\overline{\overline{u}}}_j \overline{\overline{\overline{\omega}}} + C_S(\Delta x)^4  |\overline{\overline{S}}| \frac{\partial \nabla^2 \overline{\overline{\omega}}}{\partial x_j} + 
    C_R \big(\overline{\overline{\overline{u_j'}}~ \overline{\overline{\omega'}}} - \overline{\overline{\overline{u_j'}}}~ \overline{\overline{\overline{\omega'}}} \big), \label{eq:closure_ssd}
\end{equation}
where for any $\phi$, we define $\overline{\overline{\phi'}} = \overline{\overline{\phi}} - \overline{\overline{\overline{\phi}}}$ and $\overline{\overline{\overline{\phi'}}} = \overline{\overline{\overline{\phi}}} - \overline{\overline{\overline{\overline{\phi}}}}$. Both LHS and RHS in Eq. \eqref{eq:closure_ssd} can be computed given only the resolved fields ($\overline{u}_j, \overline{\omega}$). Thus, assuming that the optimal free parameters $C_S$ and $C_R$ are equal in Eqs. \eqref{eq:closure} and \eqref{eq:closure_ssd}, we estimate them from Eq. \eqref{eq:closure_ssd} using the least squares minimization
\begin{gather}
    C_S = \frac{\langle (l_j - h_j) m_j \rangle}{\langle m_j m_j \rangle}, \label{eq:C_S} \\
    C_R = \frac{\langle (l_j-h_j-C_S m_j) b_j \rangle}{\langle b_j b_j \rangle}, \label{eq:C_R}
\end{gather}
where $l_j = \overline{\overline{u}_j \overline{\omega}} - \overline{\overline{u}}_j \overline{\overline{\omega}}$, $h_j=\overline{\overline{\overline{u}}_j \overline{\overline{\omega}}} - \overline{\overline{\overline{u}}}_j \overline{\overline{\overline{\omega}}}$, $m_j=(\Delta x)^4  |\overline{\overline{S}}| \frac{\partial \nabla^2 \overline{\overline{\omega}}}{\partial x_j}$, $b_j=\overline{\overline{\overline{u_j'}}~ \overline{\overline{\omega'}}} - \overline{\overline{\overline{u_j'}}}~ \overline{\overline{\overline{\omega'}}}$, and $\langle \cdot \rangle$ denotes horizontal averaging. Equations \eqref{eq:C_S}-\eqref{eq:C_R} predict unique coefficients $C_S$ and $C_R$ for every fluid layer and time step. Note that we first estimate the Smagorinsky coefficient $C_S$ (Eq. \eqref{eq:C_S}), and only then the backscattering coefficient $C_R$ (Eq. \eqref{eq:C_R}) for better numerical stability  \cite[see][appendix D2]{perezhogin2023subgrid}.

\subsubsection{Numerical implementation}
We consider the 1D filter of \cite{sagaut1999discrete}
\begin{equation}
    \overline{\phi}_i^x = \frac{\epsilon^2}{24}(\phi_{i-1}+\phi_{i+1}) + (1-\frac{\epsilon^2}{12})\phi_i,
\end{equation}
which has $\mathrm{FGR}=\overline{\Delta}/\Delta x=\epsilon \leq \sqrt{6}$. We apply it along two horizontal directions to create a 2D LES filter $\overline{(\cdot)} \equiv \overline{\overline{(\cdot)}^y}^x = \overline{\overline{(\cdot)}^x}^y$. An additional filter with $\mathrm{FGR}=\sqrt{12}$ is given by two iterations of 2D filter with $\epsilon=\sqrt{6}$. The boundary values are set to zero for every iteration of the 2D filter. Changing boundary conditions to $\partial_y u =0$ and $\partial_x v =0$ when the filter is applied to the velocity field may be favorable to eliminate commutation error with the curl operator near the boundary (not yet implemented in MOM6). The commutation error away from the boundaries is second order in the filter width \citep{ghosal1995basic} and related to the grid spacing, which is uniform in spherical coordinates but nonuniform in Cartesian coordinates.

The inverse direction of the energy cascade in 2D fluids can result in the prediction of a negative Smagorinsky coefficient, which may potentially lead to numerical instabilities. This issue is often resolved in 2D fluids by clipping the numerator of Eq. \eqref{eq:C_S} before spatial averaging, which is an ad hoc approach. 
In this work, we clip Eq. \eqref{eq:C_S} only after the spatial averaging. Our clipping is activated occasionally during a transient startup phase, in the lower fluid layer, and at the coarsest resolutions. Further, the coefficient $C_S$, as predicted by Eq. \eqref{eq:C_S}, is passed to the eddy viscosity module of MOM6, which is implemented using momentum fluxes for stability reasons. We ensure that $C_S$ is mostly positive and has physically relevant values ($C_S \approx 0.05$) by using an eddy viscosity model that has a positive correlation with subfilter fluxes (biharmonic Smagorinsky model) and by increasing the $\mathrm{FGR}$ parameter. 

We found that $C_R$, as predicted by Eq. \eqref{eq:C_R}, is sensitive to the artifacts near the boundary. Therefore, we exclude the ten points closest to the boundary from the horizontal averaging operator $\langle \cdot \rangle$, achieving accurate predictions ($C_R\approx 20$). The exclusion of boundary points may be necessary due to the fact that horizontal averaging is performed along nonhomogeneous directions ($x,y$). This issue might be mitigated by employing a localized formulation of the dynamic procedure using Lagrangian averaging \citep{meneveau1996lagrangian}.



 
\subsection{Thickness closure} \label{sec:thickness_closure}
We propose a new subfilter closure for thickness fluxes following \cite{bardina1980improved} as
\begin{equation}
    \lambda_j \approx \overline{\overline{u}_j \overline{h}} - \overline{\overline{u}}_j \overline{\overline{h}}.
\end{equation}
The predicted thickness flux is converted to a streamfunction $\Psi_j$ defined on the interfaces (upper and lower) of the fluid layer similar to  \cite{gent1990isopycnal} as follows: $\Psi_j|_{\text{upper}} - \Psi_j|_{\text{lower}} = \lambda_j$ with the boundary conditions $\Psi_j=0$ at the ocean surface and 
 floor. Note that such conversion is not unique and may be challenging in vanishing fluid layers. Our implementation is numerically stable as a result of three factors: using a dissipative numerical scheme for thickness equation, converting thickness fluxes to a streamfunction, and limiting the streamfunction to prevent negative thickness.

\section{Results}


Our analysis starts by evaluating the vorticity closure (Section \ref{sec:vorticity_closure}) in the double-gyre configuration of the MOM6 ocean model. We examine three filter widths $\overline{\Delta}$ (in relation to the Rossby radius), and subsequently consider three values of the sole input parameter to the subfilter closure, $\mathrm{FGR}=\overline{\Delta}/\Delta x$.




\begin{figure}
\centering{\includegraphics[width=1\textwidth]{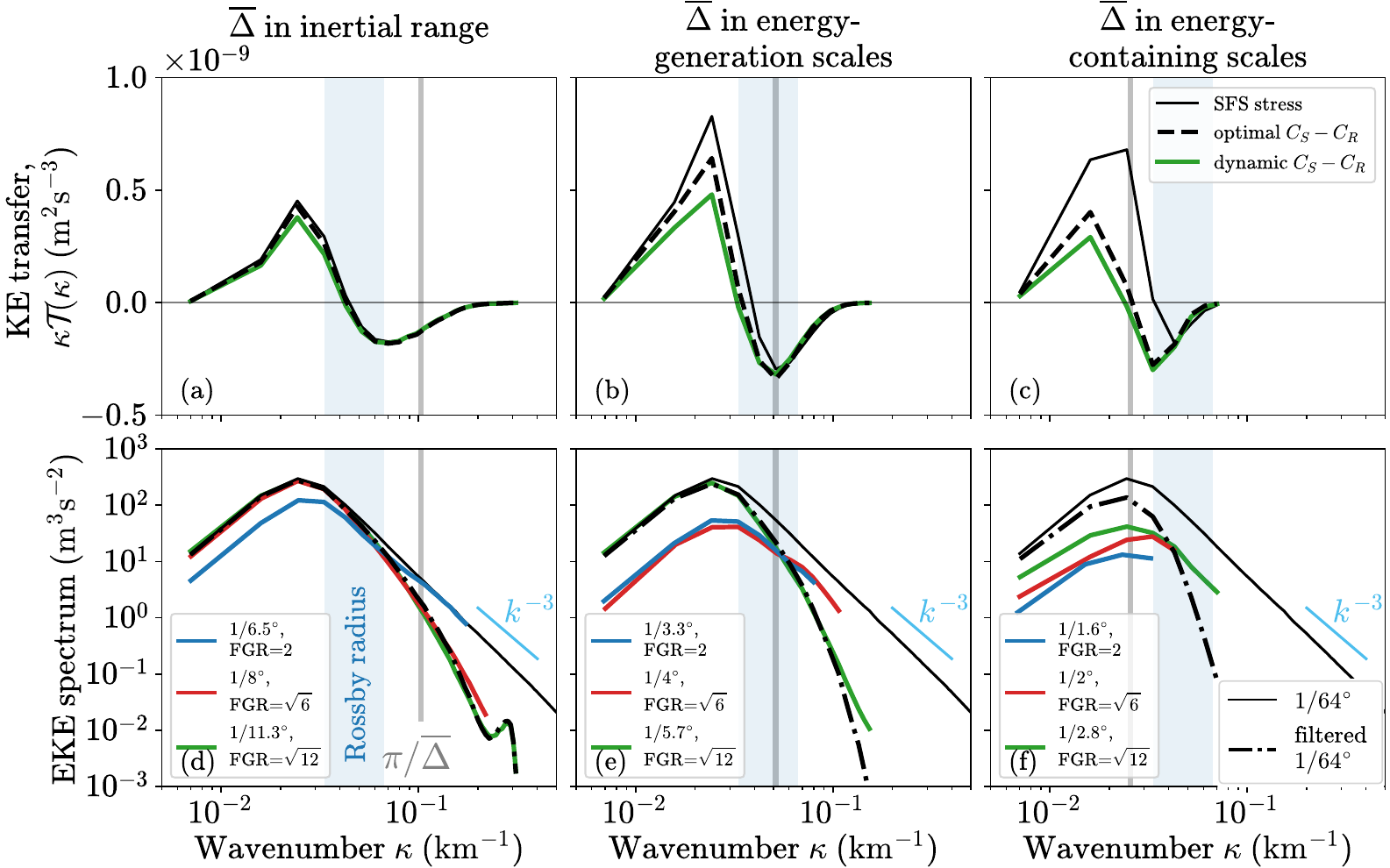}}
\caption{Subfilter KE transfer in \emph{a-priori} analysis at $\mathrm{FGR}=\sqrt{12}$ (a-c) and spatial spectrum in \emph{a-posteriori} simulations at three FGRs (d-f) for vorticity closure described in Section \ref{sec:vorticity_closure}. The backscatter can be seen as the positive energy transfer in panels (a-c). All spectra correspond to a box $(5^\circ$$-$$15^\circ\mathrm{E})$$ ~\times~$$(35^\circ$$-$$45^\circ\mathrm{N})$ in the upper fluid layer. Each column shows a different filter scale $\overline{\Delta}$ denoted with a vertical gray line. The dashed line in panels (a-c) shows vorticity closure with parameters $C_S$ and $C_R$ inferred from \emph{a-priori} subfilter flux data $\sigma_j$, which is not available \emph{a-posteriori}. }
\label{Fig2}
\end{figure}

\subsection{A-priori accuracy}
The main challenge in parameterizing vorticity fluxes is the need to predict strong KE backscatter, which appears as a prominent positive energy transfer in large scales (Figure \ref{Fig2}(a-c)).
The proposed vorticity closure accurately predicts the energy transfer spectrum \emph{a-priori} when the filter scale $\overline{\Delta}$ corresponds to the inertial range of the energy spectrum (see green line in Figure \ref{Fig2}(a)). However, when the filter scale is outside of the self-similar range, the backscatter prediction is too small compared to the subfilter data (see Figure \ref{Fig2}(b,c)). 

The drop in \emph{a-priori} accuracy can be explained in two alternative ways: (1) The desired energy transfer cannot be expressed with the subfilter closure (Eq. \eqref{eq:closure}) for any combination of parameters $C_S$ and $C_R$ or (2) dynamic estimation of these parameters is inaccurate when the dynamic procedure is applied outside of the self-similar range. To determine which explanation applies, we estimate the optimal values of parameters $C_S$ and $C_R$ by minimizing the mean squared error in the prediction of the subfilter fluxes $\sigma_j$ diagnosed from the high-resolution simulation. A subfilter closure with optimal parameters is similar to the closure with dynamically estimated parameters (compare dashed and green lines in Figure \ref{Fig2}(a-c)). Thus, we infer that the drop in \emph{a-priori} accuracy is perhaps due to the first reason. 
From a physical perspective, subfilter modeling becomes particularly difficult because backscatter occurs already at a filter scale (see Figure \ref{Fig2}(c)). 

\begin{figure}
\centering{\includegraphics[width=1\textwidth]{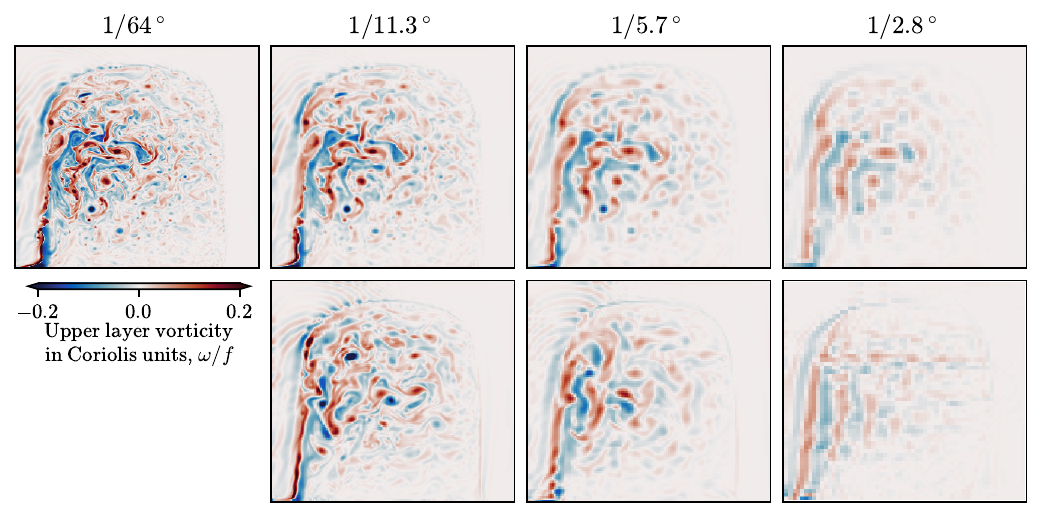}}
\caption{Snapshots of the high-resolution model and its filtered versions in the upper row and \emph{a-posteriori} simulations with the vorticity closure (Section 2.3) in the lower row; $\mathrm{FGR}=\sqrt{12}$.}
\label{Fig2_half}
\end{figure}

\subsection{A-posteriori accuracy}
We report the sensitivity of the \emph{a-posteriori} results to the $\mathrm{FGR}=\overline{\Delta}/\Delta x$ parameter.
The energy density in the small scales is too large at $\mathrm{FGR}=2$ compared to the filtered high-resolution data (Figure  \ref{Fig2}(d)) mainly because the predicted Smagorinsky coefficient ($C_S \approx 0.00-0.01$) is too small compared to a default value of $C_S=0.06$. We alleviate this issue by increasing the FGR parameter, which is possible because the dynamically estimated Smagorinsky coefficient scales as $C_S \sim \overline{\Delta}^4$. 

Panels (d) and (e) in Figure  \ref{Fig2} demonstrate the success of the explicit filtering approach---the \emph{a-posteriori} LES simulations converge to the filtered high-resolution data when the FGR parameter is increased ($\overline{\Delta}/\Delta x \to \infty$) at a fixed filter scale ($\overline{\Delta}=\mathrm{const}$). However, the rate of convergence of \emph{a-posteriori} simulations depends on the filter scale. When the filter scale corresponds to the inertial range, convergence is achieved at $\mathrm{FGR}=\sqrt{6}$ (Figure \ref{Fig2}(d)); when the filter scale corresponds to the energy-generation scales (Rossby radius), the convergence requires wider base filter with $\mathrm{FGR}=\sqrt{12}$ (Figure \ref{Fig2}(e)). Finally, at the largest filter scale, convergence to the filtered high-resolution data was not achieved (Figure \ref{Fig2}(f)). For snapshots of the solution, see Figure \ref{Fig2_half}. 

Here, we emphasize that the grid resolutions, at which the convergence to the filtered high-resolution data was observed ($\approx 1/6^\circ-1/8^\circ$), are enough to simulate the large scales directly (see Figure \ref{Fig1}(b)). This may lead to a conclusion that the vorticity closure provides only marginal improvements over the biharmonic Smagorinsky model. 
However, in the remainder of this report, we show that the proposed closure has a favorable impact on additional metrics of the mean flow even when using a large FGR or an FGR that is not sufficient to converge to the filtered high-resolution data. 

\subsection{Comparison to baseline parameterizations}

\begin{figure}
\centering{\includegraphics[width=1\textwidth]{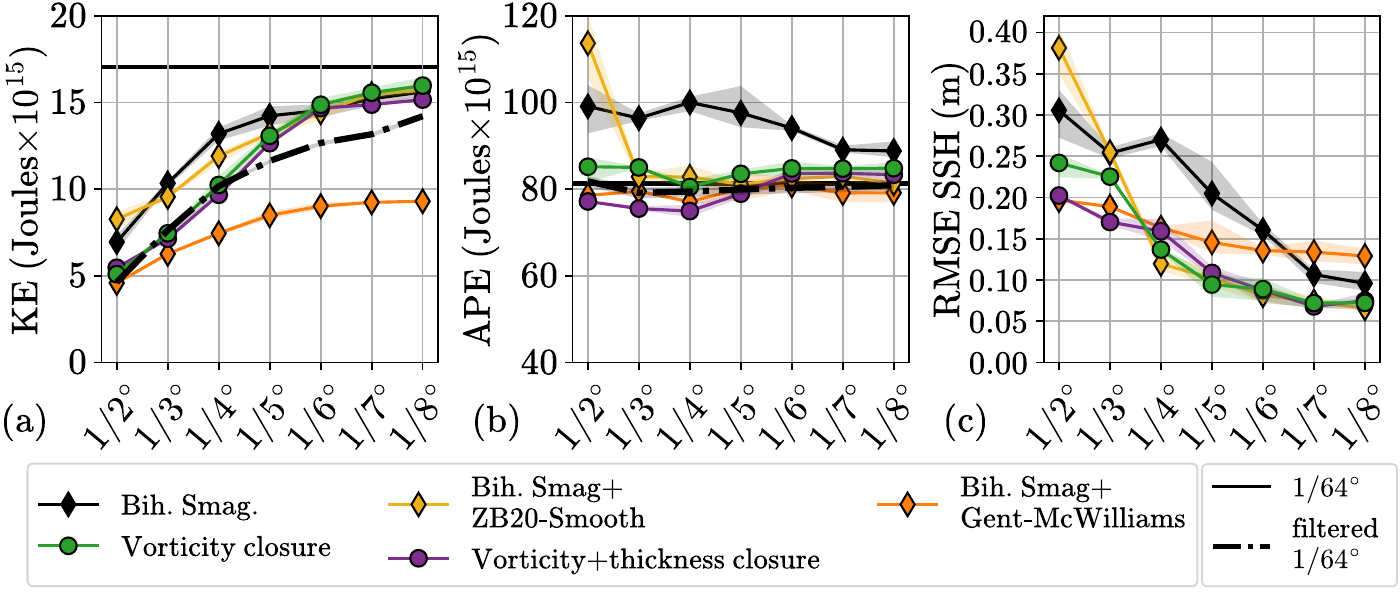}}
\caption{Three \emph{a-posteriori} metrics as a function of resolution: time-mean (a) KE and (b) APE, (c) root-mean-square error (RMSE) in time-mean SSH. Baseline parameterizations: biharmonic Smagorinsky model ($C_S=0.06$), and its combination with ZB20-Smooth \citep{perezhogin2023implementation} or \cite{gent1990isopycnal} (thickness diffusivity $125 \, \mathrm{m^2s^{-1}}$). The proposed vorticity (Section \ref{sec:vorticity_closure}) and combined vorticity+thickness (Sections \ref{sec:vorticity_closure}--\ref{sec:thickness_closure}) closures have $\mathrm{FGR}=\sqrt{12}$.} 
\label{Fig3}
\end{figure}

Based on the analysis above, we set the FGR parameter to $\sqrt{12}$, which is optimal for the intermediate grid resolutions $\approx 1/4^\circ$, and run simulations for a range of resolutions ($1/2^\circ-1/8^\circ$; see Figure \ref{Fig3}). Our baseline closure, the biharmonic Smagorinsky model, has too much available potential energy (APE) and a significant error in time-mean sea surface height (SSH). 
The proposed vorticity closure reduces the APE toward the reference value and improves the error in time-mean SSH at all resolutions (Figure \ref{Fig3}(b,c)). 


We consider another baseline: the momentum closure that was significantly tuned with respect to \emph{a-posteriori} metrics, a modification of the gradient model of \cite{zanna2020data}, abbreviated as ZB20-Smooth \citep{perezhogin2023implementation}. The proposed vorticity closure offers similar performance in all three metrics (KE, APE, error in SSH) as compared to ZB20-Smooth at resolutions $1/5^\circ-1/8^\circ$ (Figure \ref{Fig3}) but with the advantage of a more transparent tuning procedure. At the coarsest resolution ($1/2^\circ$), the vorticity closure is better in predicting the APE and SSH, although this improvement can be partially attributed to a small predicted Smagorinsky coefficient at this resolution ($C_S \approx 0.00-0.01$). Note that the vorticity closure slightly overestimates the KE of the filtered high-resolution model at resolutions $1/5^\circ-1/8^\circ$ (Figure \ref{Fig3}(a)).


Further, combining the vorticity closure with the thickness closure reduces the error in SSH at the coarsest resolutions ($1/2^\circ-1/3^\circ$; Figure \ref{Fig3}(c)). Overall, a combined vorticity+thickness closure has a performance comparable to a baseline thickness downgradient parameterization of \cite{gent1990isopycnal} at the coarsest resolutions ($1/2^\circ-1/3^\circ$) and to the momentum closure of \cite{zanna2020data} at the highest resolutions ($1/5^\circ-1/8^\circ$). Additionally, compared to the \cite{gent1990isopycnal} parameterization, the proposed thickness closure does not spuriously reduce the KE and does not require tuning of the free parameter as a function of resolution \citep[i.e., it is scale-aware; see][]{bachman2017scale}.


\section{Conclusions}
The primary difference between the large-scale ocean dynamics and 3D turbulence is the underlying flow but not the methods that can be used for their analysis and parameterization. 
In this context, we parameterize mesoscale eddies in the GFDL MOM6 ocean model with a lateral subfilter closure of vorticity fluxes by \cite{perezhogin2023subgrid} and improve it with the dynamic procedure of \cite{yuan2022dynamic} using explicit filtering. \\ 


\noindent
The successful application of the explicit filtering approach was verified by establishing convergence of the LES ocean simulations to the filtered high-resolution ocean model for sufficiently resolved grids ($1/6^{\circ} - 1/8^{\circ}$). We found that despite the common practice of injecting the KE while invoking subfilter closures (to capture backscatter),  this is not necessary in our configuration. This is primarily because the KE of the filtered high-resolution data is already small compared to the unfiltered high-resolution data at a practical FGR ($\sqrt{12}$).
\\




\noindent
We show that the proposed vorticity closure is accurate in predicting the potential energy and time-mean SSH across a range of resolutions ($1/2^\circ-1/8^\circ$), achieving similar or better results than a recently suggested parameterization \citep{perezhogin2023implementation}. We also report the limitations in the proposed model when the filter scale approaches the energy-containing scales (the \emph{a-priori} performance worsens and \emph{a-posteriori} results do not converge to the filtered high-resolution data). These issues may stem from the need to use a large FGR at coarse resolution, which increases the spatial extent of the subfilter closure. Future work should focus on developing more localized in-space closures that can accurately capture subfilter fluxes at coarse resolution. \\

\noindent
Finally, we show that the LES methods can be successfully applied for the prediction of thickness fluxes and propose a model of \cite{bardina1980improved} for this purpose. It further improves the mean flow at the coarsest resolutions ($1/2^\circ-1/3^\circ$) without spuriously reducing the KE and thus may be a good alternative to \cite{gent1990isopycnal} parameterization.

\subsection*{Acknowledgments}
\noindent Support was provided by Schmidt Sciences, LLC. This work was supported in part through the NYU IT High Performance Computing resources, services, and staff expertise. We are grateful to Dr. Parviz Moin for his support of this research direction, and to Dr. Perry Johnson for his insightful remarks on the report. We also wish to thank Dr. Andrey Glazunov for his inspiration, which helped shape the focus of this research.



\bibliographystyle{ctr}
\bibliography{brief}

\end{document}